\documentclass[sn-mathphys,Numbered]{sn-jnl}

\usepackage{graphicx}
\usepackage{multirow}
\usepackage{amsmath,amssymb,amsfonts}
\usepackage{amsthm}
\usepackage{mathrsfs}
\usepackage[title]{appendix}
\usepackage{xcolor}
\usepackage{textcomp}
\usepackage{manyfoot}
\usepackage{booktabs}
\usepackage{algorithm}
\usepackage{algorithmicx}
\usepackage{algpseudocode}
\usepackage{listings}

\let\orcidlogo\relax
\usepackage{orcidlink}

\newcommand{\cnum}[1]{%
  \raisebox{0.1ex}{\textcircled{\raisebox{-0.2ex}{\small #1}}}%
}

\newcommand{\cnumlabel}[1]{%
  \raisebox{0.1ex}{\textcircled{\raisebox{-0.35ex}{\small #1}}}%
}

\usepackage{lineno}
\usepackage[normalem]{ulem}

\begin{document}

\historywhiteon

\received{xx xx xxxx} 
% \revised{xx xx xxxx} 
\accepted{xx xx xxxx}

\title[Article Title]{Autonomous thermodynamic cycles via robotic mobility and sensing
}

\author*[1]{\fnm{Sofia} \sur{Kuperman \orcidlink{0000-0003-2710-6019}}}\email{sofia.kup@campus.technion.ac.il}

\author[1]{\fnm{Ezra} \sur{Ben-Abu \orcidlink{0000-0001-7252-2965}}}
\author[1]{\fnm{Yaron} \sur{Veksler \orcidlink{0000-0002-3043-5940}}}
\author[1]{\fnm{Anna} \sur{Zigelman \orcidlink{0000-0001-7597-8773}}}
\author[1]{\fnm{Sefi} \sur{Givli \orcidlink{0000-0003-3329-8731}}}
\author[1]{\fnm{Amir D.} \sur{Gat \orcidlink{0000-0002-1753-8937}}}

\affil[1]{\orgdiv{Faculty of Mechanical Engineering}, \orgname{Technion – Israel Institute of Technology}, \orgaddress{ \city{Haifa}, \postcode{3200003}, \country{Israel}}}

\abstract{
Thermodynamic cycles are the foundation of energy conversion across natural and engineered systems, transforming heat into useful work. However, these cycles traditionally operate between fixed thermal reservoirs, restricting them to specific locations and temperature differences. Here, we introduce autonomous thermodynamic cycles enabled by robotic mobility and sensing, allowing robots to perform thermodynamic cycles by accessing spatially varying temperature fields. We experimentally realize this concept using multistable gas-filled capsules that circulate within the system across a thermal gradient. Our model reveals that rapid transitions in the capsules’ energy states allow the system to operate as a mobile heat engine that harvests and stores energy. By linking the capsule-scale internal energy dynamics to the robot’s large-scale navigation strategy, we optimize locomotion paths that balance motion cost and energy harvesting. These findings demonstrate that thermodynamic cycles can emerge when autonomous systems navigate their environments, offering an artificial analog of organisms that forage for energy across spatial resources.
}


\maketitle

\section*{Introduction}\label{Introduction section}

Thermodynamic cycles have been the cornerstone of energy conversion for over two centuries, transforming thermal gradients into useful work \cite{kitanovski2025emerging,tchanche2011low}. From the Rankine cycle to modern combined cycles, these processes rely on heat exchange between reservoirs at different temperatures to generate power \cite{woudstra2010thermodynamic,chen2010review}. However, their dependence on stationary thermal reservoirs anchors their operation to locations where such reservoirs exist \cite{beer2007high,lecompte2015review}. 
Nature offers a different strategy, where ectotherms, or cold-blooded animals, sustain themselves by moving between warmer and cooler regions in their environment to maintain functional body temperatures \cite{ectotherm1989evolution,angilletta2002evolution}. Translating this principle to engineered systems suggests thermodynamic cycles that actively seek favorable thermal environments, requiring perception to map thermal gradients, planning to select favorable regions, and mobility to reach them.

Thermodynamic cycles enable energy harvesting across diverse conditions, even at low-grade thermal differences \cite{lee2014electrochemical,yang2014charging}. Such differences are commonly found in nature, typically spanning 10-50 °C across meter scales, and arise from variations in material properties, light exposure, phase, or diurnal cycles \cite{cottrill2018ultra,kishore2019energy}. For instance, asphalt under summer sun can exceed 65 °C \cite{asphalt2017energy} while adjacent grass remains around 25 °C \cite{shade_grass2012effect}, and deserts routinely fluctuate from over 55 °C daytime highs to sub-zero nights \cite{desert2023temperature}. Such gradients represent distributed thermal resources that could potentially be exploited by mobile systems.

Autonomous mobile robots embody the sensing, planning, and mobility capabilities required for thermodynamic systems to exploit such thermal landscapes. By navigating between regions of different temperatures, mobility can effectively replace stationary thermal reservoirs, allowing thermodynamic cycles to operate using environmental thermal resources. The resulting energy can supply power for robotic operation and alleviate endurance limitations, a primary bottleneck in applications where long-duration deployment is essential, such as environmental monitoring, disaster response, and planetary exploration \cite{dunbabin2012robots,ecorobotics_mazzolai2025environmental}. In such scenarios, reliance on charging infrastructure or battery replacement is often infeasible, limiting mission duration and operational autonomy \cite{Biomimetic_energy_storage_in_robots2025Pikul,wyder2025robot,ross2016controlling,animals_outrun_robots2024}. Since energy harvesting depends on environmental temperature fields, locomotion itself becomes part of the energy strategy. Navigation can therefore favor trajectories that pass through thermally advantageous regions rather than strictly minimizing path length. Similar approaches appear in bird migration, where long-distance journeys incorporate detours through resource-rich regions that support energy replenishment \cite{alerstam1990optimal,aikens2024learning}.

To address the endurance limitations, efforts to extend the operational range and duration of autonomous systems have included pairing electrochemical storage with passive energy harvesting \cite{energy_harvesting_storage_robots2023,hassan2020renewable,mariani2026soft}, using photovoltaics \cite{thomas2006energy,solar_hand_launchable_UAV2015,solar2016}, kinetic energy recovery \cite{Flight_Sailing_modes2024Mirco,Wave_Powered_UAV2009}, and thermoelectric harvesting from residual heat \cite{liang2023next}. A complementary strategy  is to couple energy acquisition with robotic locomotion and decision-making, allowing systems to actively exploit environmental energy resources in ways that are programmable, scalable, and environmentally safe.

Realizing this vision requires a physical medium capable of supporting reconfigurable and distributed architectures within the robot’s body \cite{alu2025roadmap}. Here, we employ metafluids as the underlying medium, as they provide a foundation for realizing such embedded energy systems. Their programmable properties have been demonstrated across domains such as viscosity modulation, optical tuning, and energy-related applications \cite{Bertoldi_programmable_metafluids_2024,peretz2022metafluid,peretz2023multistable,Metamaterial_robotics2025}. Within the emerging class of embedded energy networks \cite{Fish2019electrolytic_blood,embodied_energy2022,Jellyfish2024}, which rely on liquid carriers, we instead use deformable energy-bearing capsules that operate reversibly without requiring replenishment, analogous to circulating blood cells. These units are inherently scalable and tunable, and can be safely implemented using encapsulated gases such as air.

In this work, we introduce an internal distributed energy system composed of elastically multistable, gas-filled capsules. Each capsule, with incorporated magnetic edges, moves within coil-wrapped tubes, enabling efficient energy transduction during rapid transitions between elastic stable states using electromagnets. This architecture enables cyclic harvesting and storage of environmental thermal energy and releases it on demand. The central novelty of our approach is embedding thermodynamic cycles within a distributed internal energy architecture that harvests environmental thermal energy through mobility. We analyze the coupled dynamics and thermodynamics of the system to establish design principles for operation in diverse thermal fields. The resulting framework provides a programmable, scalable, and environmentally safe strategy for robotic energy management,  capable of  supporting sustained autonomy by drawing power from environmental thermal gradients. Initial experimental demonstrations validate the concept and show how such systems can generate power using thermodynamic cycles, opening new directions for long-duration, actively foraging robotic autonomy.

\section*{Results}\label{Results}

\subsection*{Embedded energy via multistable unit cells}\label{Subsection Fig 1 main text}
 
We present a framework for performing thermodynamic cycles that harvest and store energy from temperature variations. These processes are enabled using gas-filled multistable capsules distributed within an internal network. In this approach, thermal changes and external loads can expand or collapse the capsules, potentially triggering transitions between their stable states. These transitions result in rapid volume changes that correspond to discrete energy levels associated with local energy minima.

\begin{figure}[h!]
\centerline{\includegraphics[width=1\textwidth]{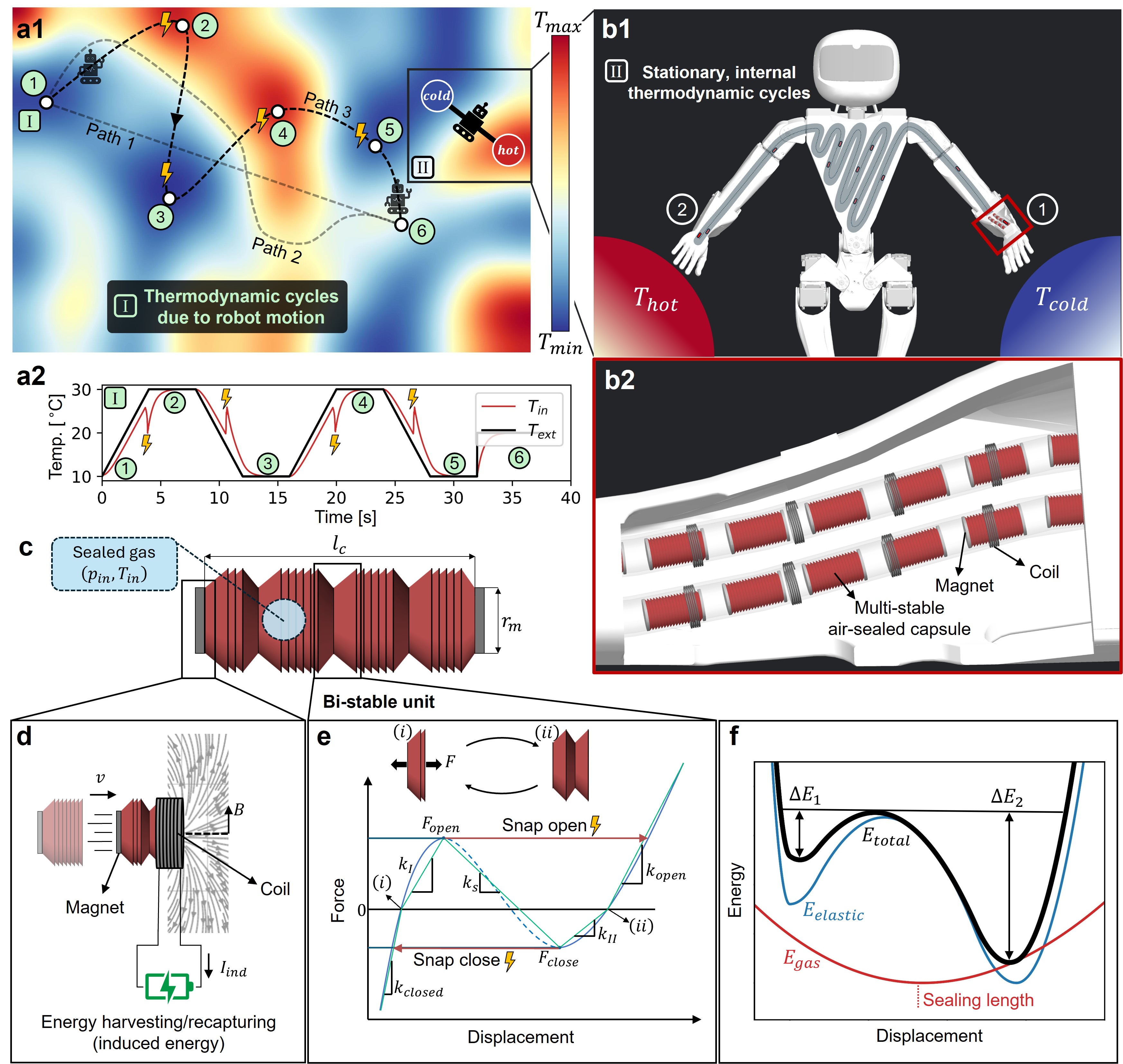}}
{\caption{\textbf{Energy storage and harvesting mechanisms using gas-filled multistable capsules.} (a1) Robot paths across a spatially varying thermal environment, illustrating two operating modes: (I) energy harvesting by traversing alternating hot and cold regions and (II) stationary operation using local temperature gradients. (a2) Corresponding internal capsule temperature along a representative trajectory (Path 3), showing thermal lag and state-transition events (marked by lightning symbols), where $T_{in}$ and $T_{ext}$ denote are the capsule’s internal and external temperatures, respectively. (b1) Schematic of embedded capsules as distributed energy units within a robotic platform. (b2) Capsules that are transported within an internal vascular system, actuated by solenoids. (c) The sealed gas-filled multistable capsule composed of a series of bistable units with magnets at its edges. (d) Electromagnetic energy conversion during rapid capsule transition, induced by the motion of a mounted magnet relative to a surrounding coil. (e) Piecewise-linear approximation of the force-displacement response of a bistable unit, highlighting two stable states and snap force thresholds ($F_{open}$, $F_{close}$). (f) Effective energy profile of a bistable capsule, combining elastic and gas-pressure contributions, and the energy barriers $\Delta E_1$ and $\Delta E_2$ between closed and open states, respectively.}
\label{Fig 1 - Schematic}}
\end{figure}

The ability to perform thermodynamic cycles depends on the robot's route through spatially varying thermal fields, which affect the internal temperature of the embedded capsules and whether they undergo the state transitions required for thermodynamic operation. A robot traveling from an origin to a target can follow multiple trajectories that sample different ambient temperatures (see Fig. \ref{Fig 1 - Schematic}a1-2). Consequently, the net energy yield is path-dependent. While the direct trajectory (Path 1) or a deviation that remains below the transition threshold (Path 2) completes the task without generating power, a strategically chosen trajectory (Path 3) traverses thermal gradients that exceed the threshold, triggering transitions that result in energy harvesting (indicated by lightning symbols). 

We classify possible energy cycles into two modes: (I) mobile cycles, in which the robot encounters temperature variations during motion, and (II) stationary cycles, where capsules circulate within the robot between regions in contact with hot and cold environments (see schematic in Fig.~\ref{Fig 1 - Schematic}b1). Stationary cycles can be performed where local temperature gradients naturally arise from heterogeneity in nearby materials, phases, or heat sources. For such cycles, metafluids offer a programmable alternative to conventional working fluids, with distinct advantages including a tunable operating range using safe fluids, such as air, ambient pressure energy storage, and operation driven by both positive and negative temperature changes. To realize either mode, the energy-bearing capsules circulate within a tubular network surrounded by solenoids (Fig.~\ref{Fig 1 - Schematic}b2). Integrating magnets at the capsule ends enables electromagnetic interaction with the solenoids.

The capsules act as the system’s energy storage unit cells. Each unit cell gas-filled element composed of a series of bistable units (Fig. \ref{Fig 1 - Schematic}c). During rapid transitions, the kinetic energy is converted into electrical energy via induction via the surrounding solenoids (Fig. \ref{Fig 1 - Schematic}d), rather than being dissipated as loss (for additional details, see SI \S2.3.3). These transitions, called snap-through events, occur when  a force applied on the element is beyond the threshold forces $F_{close}$ or $F_{open}$ required for corresponding state transition, as shown in the force-displacement profile of the bistable unit in Fig. \ref{Fig 1 - Schematic}e. This unit exhibits two stable states, denoted closed and open. Once a snap-through occurs, the element remains in its new state until an opposing net force is applied. The non-linear response is captured by a penta-linear model comprising four stiffness segments ($k_{closed}, k_{I}, k_{II}, k_{open}>0$) and a spinodal region ($k_{s}<0$), where elastic resistance is momentarily inverted (see SI \S1.1 for experimental characterization), producing high edge velocities. Fig. \ref{Fig 1 - Schematic}f illustrates the effective energy profile of the capsule, which combines elastic deformation with gas contributions (see SI \S 2.5 for model details). The energy barriers between the stable states, $\Delta E_{1}$ and $\Delta E_{2}$, determine the energy storage capacity attainable through state transitions. Importantly, this energy profile can be tuned to suit specific requirements, providing a programmable mechanism for energy harvesting and on-demand actuation.
 
We examine the energy storage and harvesting behavior of these multistable capsules via two actuation modes: (i) external thermal excitation, arising from exposure to environmental temperature variations, and (ii) internal electromagnetic actuation, achieved by driving current through a solenoid. In the following section we focus on the internal actuation mode, which enables on-demand energy transitions within the capsules.

\subsection*{Equilibrium states of multistable capsules}\label{Equilibrium states exp Sec}

\begin{figure}[h!]
\centerline{\includegraphics[width=0.78\textwidth]{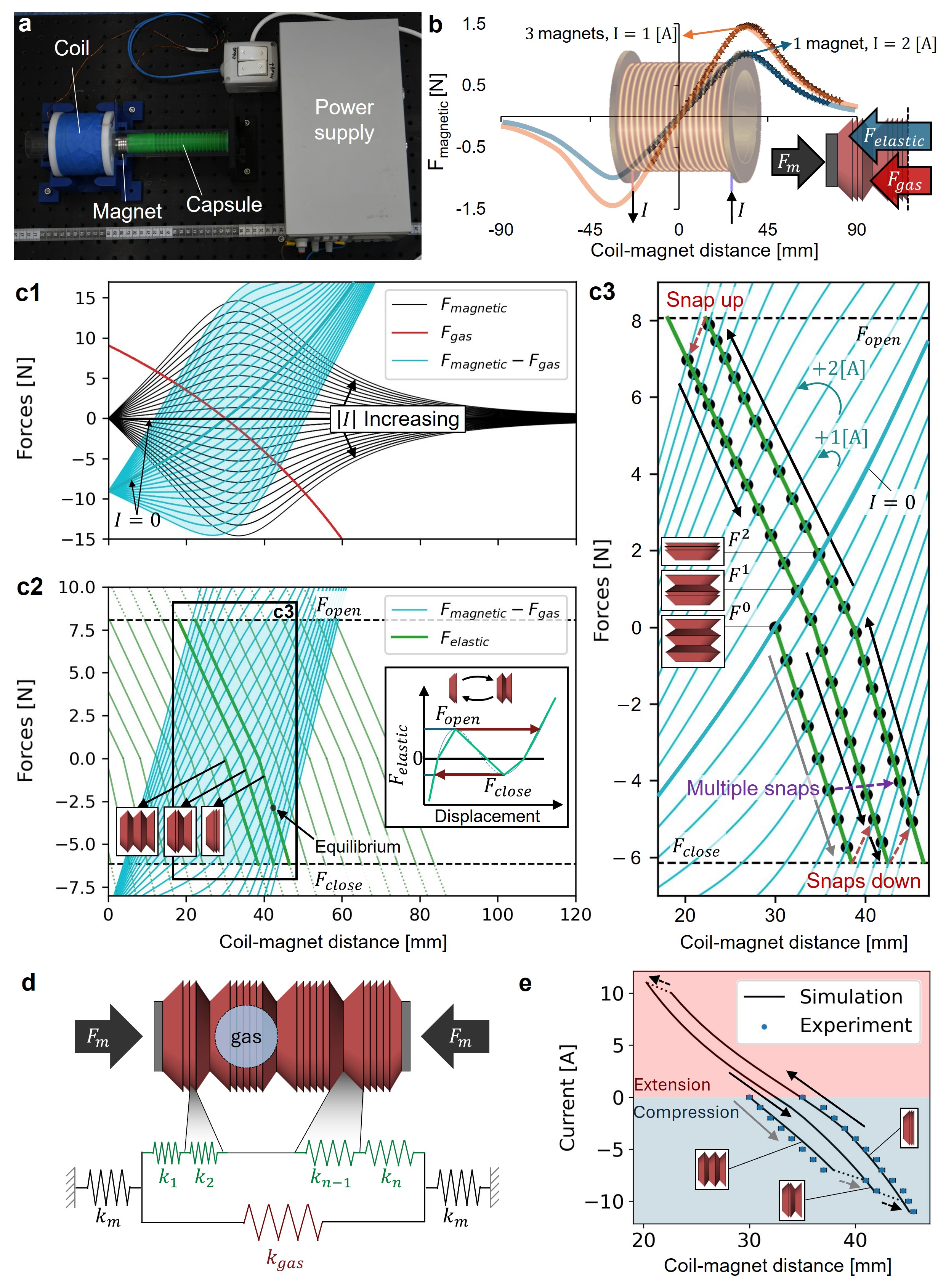}}
{\caption{\textbf{Equilibrium states of electromagnetically actuated capsules.}
(a) Experimental setup consisting of a sealed capsule with magnets at its edges, and a power supply to drive current through a coil. (b) Comparison of theoretical (solid lines) and measured (circles) electromagnetic forces as a function of magnet distance relative to the center of the solenoid for two representative solenoid currents and magnet numbers, alongside a schematic of the forces acting on the capsule. (c1) Net actuation force resulting from the applied magnetic force and opposing internal gas force. (c2)
Elastic forces corresponding to different stable states of the capsule. Equilibrium positions occur at the intersection of the net force and elastic force. Varying the net force within a given stable state shifts the equilibrium along the corresponding elastic response. The shaded region represents all possible stable equilibria. Once the force exceeds the snap-through thresholds, $F_{close}$ or $F_{open}$, instability occurs and the capsule transitions to another stable state. Transitions between elastic force responses associated with different stable states occur only through snap-through.  The inset shows the snap force thresholds $F_{close}$ and $F_{open}$. (c3) Close-up of the magnetically actuated cycles where currents are cyclically increased and decreased incrementally. Stable equilibria are indicated by filled black circles, the thick line denotes the net force in the absence of an external magnetic force. Red dashed arrows denote the unstable snap-through transitions when elastic forces exceed $F_{close}$ or $F_{open}$, and purple dashed line denotes a sudden current increase by 4 A leading to multiple snaps. Residual gas forces are represented by $F^{i}$. Gray arrow traces the initial trajectory and black arrows indicate the repeating cycle. The legend is shared between panels (c2) and (c3). (d) Effective spring model of the loaded capsule, used to determine equilibrium and stability. (e) Comparison between simulations and experiments for magnet location as a function of solenoid current. In panels (b) and (e) error bars are smaller than the marker size.}
\label{Fig 2 - Equilibrium states}}
\end{figure}

We theoretically and experimentally characterize the equilibrium response of the gas-filled multistable capsules with edge-mounted magnets under solenoid generated magnetic forces. This analysis identifies the actuation range and the conditions for state transitions that define the capsule energy states, which generalizes to other externally applied forces. We first characterize the magnetic force exerted by the solenoid on the magnet as a function of axial distance from the coil center (see SI \S1.4). Fig. \ref{Fig 2 - Equilibrium states}b compares experimental measurements with theoretical predictions \cite{robertson2012axial} for two representative configurations differing in drive current and magnet number, showing good agreement. 

The validated magnetic force is used to determine the net actuation force, given by the difference between the magnetic force and the opposing gas force, shown in Fig. \ref{Fig 2 - Equilibrium states}c1. The net force is incorporated into a model that accounts for the elastic force of the multistable structure (see SI §2.2),  from which the equilibrium states are determined. These equilibria occur at the intersections of the elastic force and the net force applied on the capsule, where force balance is satisfied $F_{\text{elastic}}=F_{\text{magnetic}}-F_{\text{gas}}$. Fig. \ref{Fig 2 - Equilibrium states}c2 shows equilibria of the capsule under coil currents sampled from -11 to 11 A in 1 A increments for visualization, although it is tunable continuously.  
State transitions are governed by the elastic response and occur as rapid snap-through events when $F_{\text{elastic}}>F_{open}$ (snap-up) or $F_{\text{elastic}}<F_{close}$ (snap-down). Although the snap onset is predictable, the transitioning element cannot be predetermined due to the similarity among the bi-stable elements in the capsule. When an element switches state, its volume and stiffness change, modifying the effective capsule stiffness, as reflected by the slopes of the elastic-force curves (SI Fig. S12). The blue region defines the full operational window of the capsule over this current span. For a given initial coil-magnet distance, it therefore bounds both the number and spatial range of possible snap events. In this example, there are three stable states within $F_{close}<F<F_{open}$, each corresponding to different energy levels.

Reversible transitions are achieved by applying negative and positive coil currents which compress or extend the capsule to induce snap-down and snap-up events. A close-up of the experimental cycles in Fig. \ref{Fig 2 - Equilibrium states}c3 shows the capsule length change progressively along stable equilibria (black circles), until the force threshold is crossed, triggering unstable and dynamic snap-through events (red dashed arrows). A single actuation step can also induce multi-state transitions (purple dashed arrow) when the applied force exceeds multiple transition thresholds. Two types of cycles are revealed, a repeatable cycle (black) and a cycle containing an initial, non-repeatable segment (gray). The latter depends on the initial magnet position and capsule state, and can be accessed only once due to the limit of the attainable net-force envelope set by the available current range. As the capsule shifts away from the sealing state, residual gas forces accumulate, creating stepwise offsets along the zero-current curve, $F^{(0)} < F^{(1)} < F^{(2)}$, where the superscript denotes the capsule state ($0$ corresponds to the sealing state).

We analyze the stability of the equilibrium sates by modeling the system as an effective spring network, and evaluate the stability criterion based on the Hessian matrix of the total potential energy (see SI \S2.2.1). Fig. \ref{Fig 2 - Equilibrium states}e compares current-controlled compression experiments and simulations obtained by solving our model and showing strong agreement. This agreement validates the framework, which readily extends to other multistable structures, coils and actuation methods, provided the relevant force-displacement relations and gas properties are implemented. With the equilibrium states characterized, we next examine the transient dynamics.

\subsection*{Dynamics and thermodynamic cycles for energy storage}\label{Subsection_dynamics_and_energy}

\begin{figure}[h!]
\centerline{\includegraphics[width=1\textwidth]{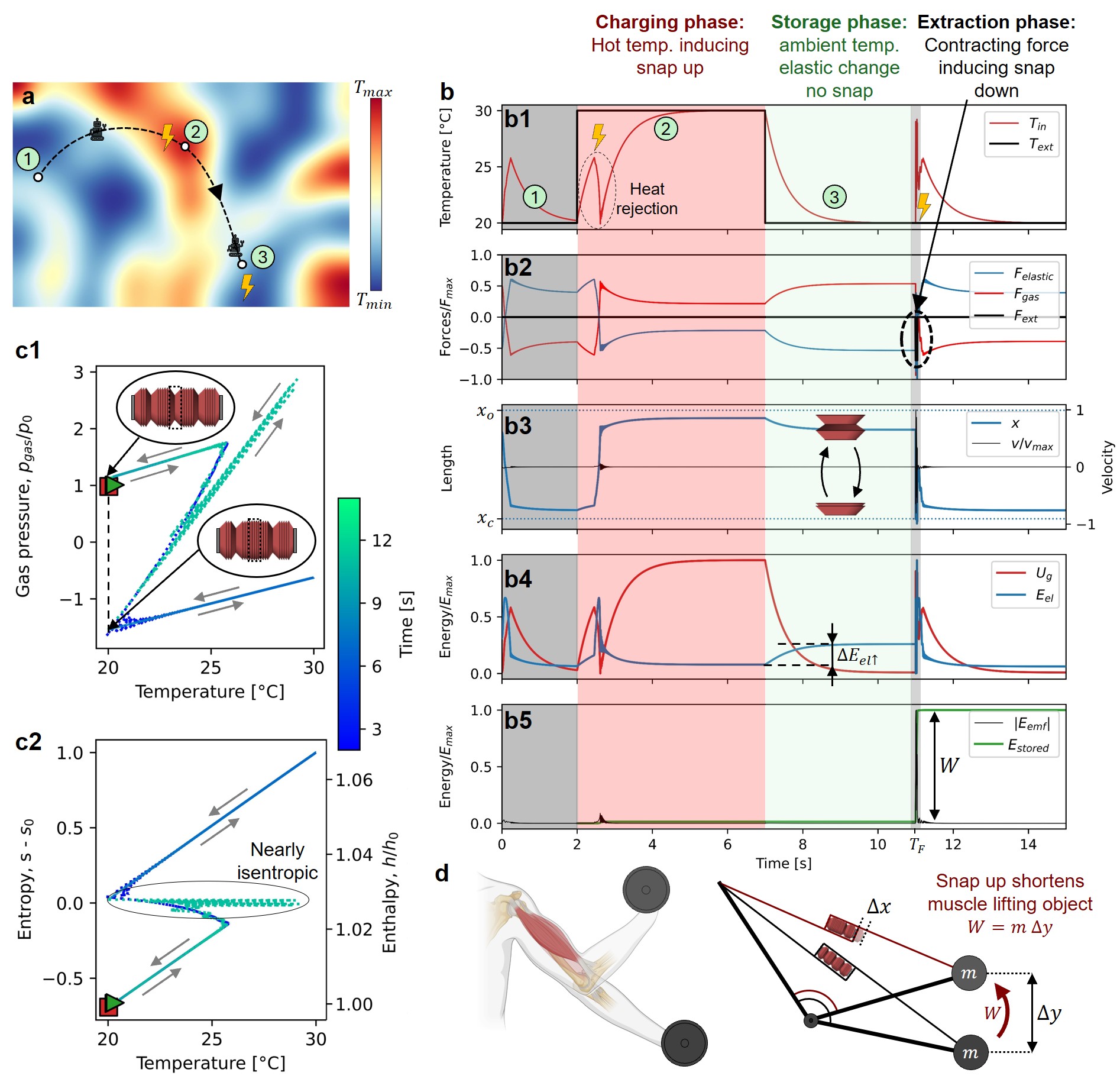}}
{\caption{\textbf{Energy storage in capsules via energetic minima and its on-demand release as mechanical work.} (a) Robot trajectory across a thermal field, illustrating energy storage through thermally driven snap-up, with subsequent energy release using electromagnetically driven snap-down. (b) Simulation of capsule dynamics: (b1) Internal and external temperatures $T_{in}$, $T_{ext}$. Expansion and snap-up upon heating, which induces heat rejection from the capsule, followed by contraction without snap-down during cooling. (b2) Force balance between elastic, $F_{\text{elastic}}$, gas, $F_{\text{gas}}$, and external force, $F_{ext}$, activated through electromagnetic force at $T_{F}$ to induce snap-down. (b3) Capsule length $x$ and velocity $v$ abruptly transitioning during snap events. (b4) The interplay between gas energy, $U_{g}$, and the elastic energy, $E_{el}$. After snap-up, cooling is insufficient to trigger snap-down, resulting in residual stored energy $\Delta E_{el\uparrow}$. (b5) The energy induced in the solenoid by the magnet motion, $E_{emf}$, the energy accumulated and retained in the system, $E_{stored}$, and the usable work, $W$. (c) The thermodynamic cycle of the capsule showing multiple stable states for the same temperature. (c1) Pressure-temperature and (c2) entropy/enthalpy-temperature diagrams showing near-isentropic snap-through behavior. The green triangle and the red square denote the start and the end states, respectively. (d) The mechanical work, $W$, used to lift a load produced by a snap-through event. 
}
\label{Fig 3 - Energy storage and work}}
\end{figure} 

To understand the capsule’s energy storage and release, we analyze its coupled dynamic and thermodynamic response under actuation. We extend the equilibrium model to incorporate inertia, heat exchange, thermodynamic relations, and electromagnetic drag and induction (full description in SI \S2). We consider a scenario in which ambient temperature induces energy shifts in the internal capsules, enabling them to store energy that can later be released as mechanical work (Fig. \ref{Fig 3 - Energy storage and work}, Movie S1). As a robot traverses a spatial temperature field, sufficiently high local temperatures trigger snap-up in the capsules (\cnum{2} in Fig. \ref{Fig 3 - Energy storage and work}a-b). During these transitions, the capsules rapidly reject heat, i.e. transfer heat to the environment, while expanding (Fig. \ref{Fig 3 - Energy storage and work}b1-3). This reveals a pronounced separation of time scales, where the capsule’s internal temperature equilibrates with the external temperature through slower heat transfer, whereas heat exchange during snap-through occurs on a much faster time scale.

If subsequent cooling is insufficient to trigger snap-down due to asymmetric transition thresholds, the capsule remains in the higher-energy configuration. This hysteresis retains the elastic potential energy, allowing the capsules to function as mechanical batteries that store elastic energy ($\Delta E_{el\uparrow}$ in Fig. \ref{Fig 3 - Energy storage and work}b4). A brief solenoid-driven electromagnetic actuation pulse can then be exerted on demand (at $T_{F}$ in Fig. \ref{Fig 3 - Energy storage and work}b) to provide the additional work needed to overcome the snap-down barrier, releasing the stored energy as useful work (Fig. \ref{Fig 3 - Energy storage and work}b5) and completing the thermodynamic cycle. 
Thermodynamic analysis shows that structural multistability gives rise to multiple thermodynamic equilibria at the same temperature, corresponding to distinct gas pressures and enthalpies, and indicates that snap-through transitions are near-isentropic (Fig. \ref{Fig 3 - Energy storage and work}c). The elastic energy stored in this process can be converted into work upon release, illustrated here by lifting a mass to a specified height (Fig. \ref{Fig 3 - Energy storage and work}d). The rapid release of stored energy as work is analogous to the role of ATP in biological muscle, which enables rapidly supplies chemical energy for muscle contraction. Together, these results demonstrate a complete thermodynamic cycle in which environmental thermal charging is separated from active discharge.

\subsection*{Energy harvesting from thermal variations}\label{Subsection Energy harvesting}

\begin{figure}[h!]
\centerline{\includegraphics[width=1\textwidth]{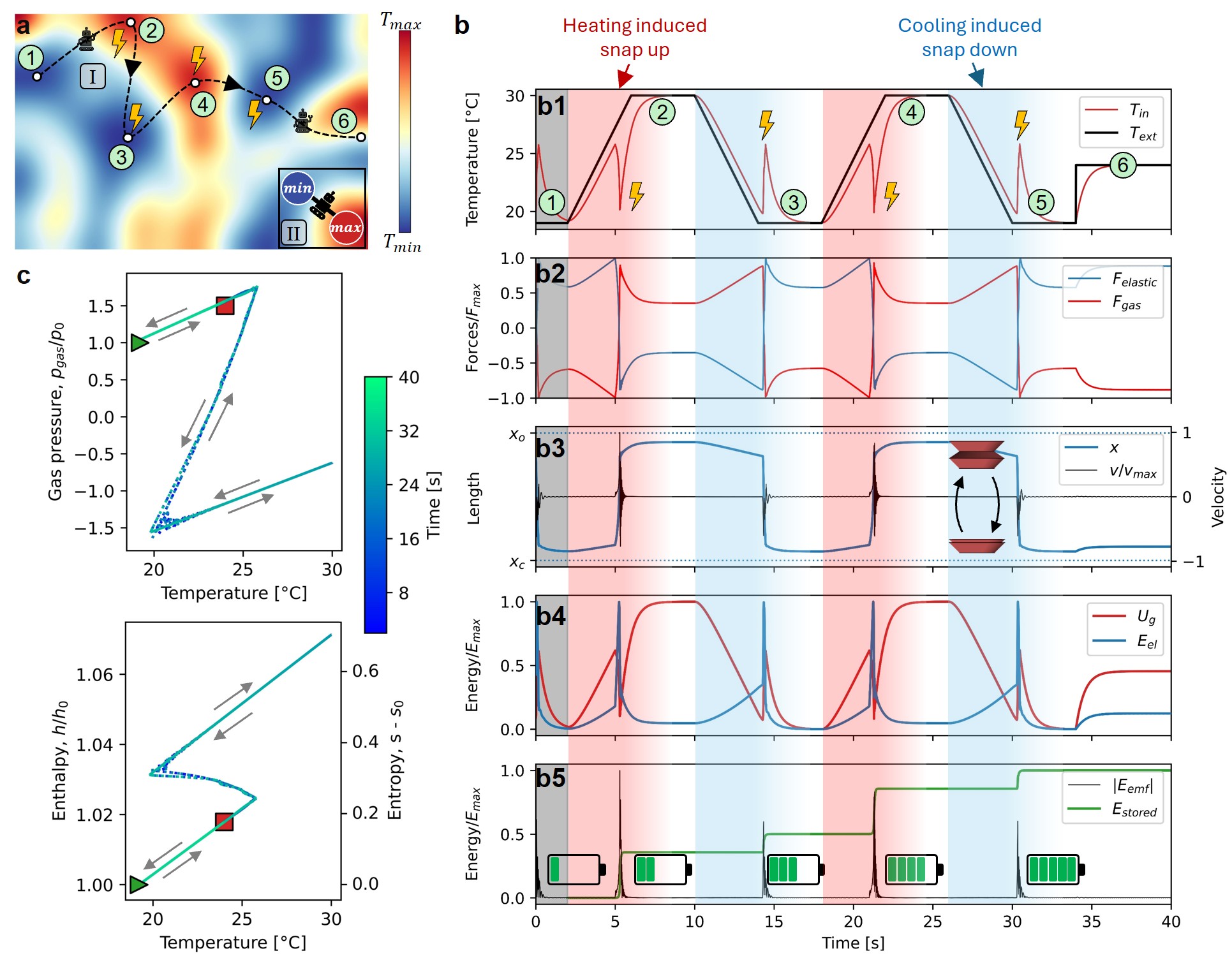}}
{\caption{\textbf{Energy harvesting from cyclic thermal variations}. (a) Energy scavenging via snap-through events through (I) a mobile system sampling temperature variations on route, or (II) a stationary system leveraging temperature gradients across the system. (b) Simulation of capsule dynamics under cyclic heating and cooling. (b1) External temperature $T_{ext}$ driving changes in the internal capsule temperature $T_{in}$ via heat transfer. (b2) Internal elastic and gas forces, $F_{\text{elastic}}$ and $F_{\text{gas}}$, interchanging according to the external temperature conditions and the capsule state. (b3) Capsule length $x$ indicating the state, and edge velocity $v$ which peaks during state transitions corresponding to high kinetic energy. (b4) Energy components: internal gas energy, $U_{g}$, and the elastic energy, $E_{el}$. (b5)  Electromagnetic energy induced as a magnet at the capsule edge moves within a solenoid, $E_{emf}$, which accumulates as stored energy, $E_{stored}$, through repeated snap-through events. (c) Capsule thermodynamics. Gas pressure, enthalpy and entropy as functions of internal temperature, showing mixed thermodynamic cycles.}
\label{Fig 4 - Energy harvesting}}
\end{figure}

Beyond energy storage, our energy system provides mechanisms for environmental energy harvesting. Ambient temperature fluctuations can trigger snap-through transitions that generate bursts of kinetic energy, which can be converted via electromagnetic induction and stored onboard (Fig. \ref{Fig 4 - Energy harvesting}, Movie S2). By selecting routes with temperature variations that trigger repeated capsule state changes, or by positioning the system across a temperature gradient, the capsules can be cyclically actuated to transition between their stable states (Fig. \ref{Fig 4 - Energy harvesting}a). As the capsules are moved between alternating heating and cooling regions, their internal temperature lags behind the external driving temperature due to the heat capacity of the encapsulated gas (Fig. \ref{Fig 4 - Energy harvesting}b1). Consequently, the gas and elastic forces alternate in dominance depending on the capsule state and volume (Fig. \ref{Fig 4 - Energy harvesting}b2). During heating and cooling the capsules exhibit gradual elastic deformation, followed by abrupt transitions during snapping events (Fig. \ref{Fig 4 - Energy harvesting}b3). These transitions generate bursts of kinetic energy associated with the high velocity of the capsule edges. Despite this rapid motion, energy is exchanged between the elastic and gas components, indicating that the capsule dynamics arise from internal energy transfer between these components (Fig. \ref{Fig 4 - Energy harvesting}b4). The kinetic energy of the magnet attached to the capsule edge during snap-through is converted via electromagnetic induction in surrounding solenoids into electrical energy, which is harvested and stored in the system with each snap event (Fig. \ref{Fig 4 - Energy harvesting}b5). In these procedures the system undergoes repeating thermodynamic cycles (Fig. \ref{Fig 4 - Energy harvesting}c). The cycle is hysteretic, and the magnitude of this hysteresis increases for capsules with higher difference in snap force thresholds (see SI \S 2.4, Fig. S16). These results demonstrate thermally driven bistable cycles that convert environmental temperature variations into harvested energy.

\begin{figure}[h!]
\centerline{\includegraphics[width=1\textwidth]{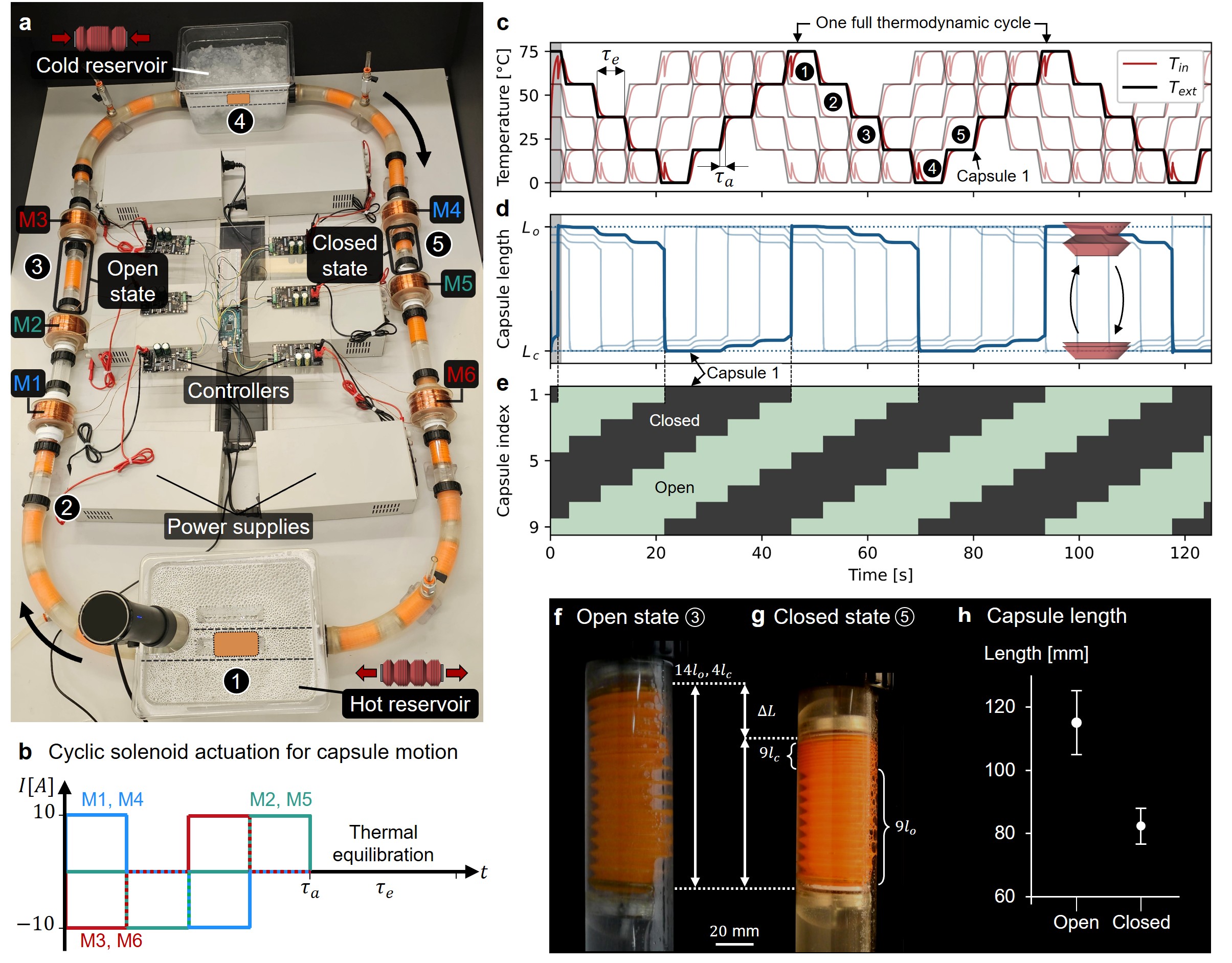}}
{\caption{\textbf{Experimental demonstration of thermally induced thermodynamic cycles.} (a) Experimental setup in which magnet-edged capsules circulate within a closed tube loop between a hot reservoir ($75^\circ$C), inducing snap-up, and a cold, ice-water reservoir ($0^\circ$C), inducing snap-down. The capsules are transported through the water-filled tube via alternating magnetic fields generated by six solenoids (M1-M6). Black rectangles  mark the locations corresponding to the state images of the capsules inpanels (f,g). (b) Repeated solenoid current waveform consisting of a pulse ($\tau_a= 110$~ms, $I\pm10$~A) that advances the capsules clockwise, followed by a dwell for thermal equilibration ($\tau_e=5$~s). (c) Tube and capsule-internal temperatures for four circulating capsules with $\tau_a+\tau_e$ phase-shift (capsule 1 shown in bold). (d) Capsules length during circulation. (e) Capsule state evolution during operation, with a nearly equal split between open and closed capsules throughout. (f,g) Representative images of open and closed capsules at \cnumlabel{3} and \cnumlabel{5}, respectively, where $l_{c}$ and $l_{o}$ denote the lengths associated with the closed and open states of the individual bistable elements comprising each capsule. (h) Mean of measured capsules' lengths associated with each state, $L_{open}=115 \pm  9.8, \, L_{closed}=82 \pm 5.6$, nominal extension $\Delta L=39.5\%$. Error bars represent one standard deviation.}
\label{Fig exp - thermal cycling}}
\end{figure}

We next demonstrate the feasibility of this concept through proof-of-concept experiments using the system shown in Fig. \ref{Fig exp - thermal cycling}a (see SI \S 1.5 for details). In this setup, capsules circulate between hot and cold reservoirs through repeated sequences of solenoid actuation. Each actuation sequence (Fig. \ref{Fig exp - thermal cycling}b) advances the capsules by one quarter of the distance between the reservoirs and provides sufficient time for thermal equilibration before the next actuation (Movie S3).
Under conduction-dominated heat transfer, the temperature along the tube changes linearly between the hot and cold reservoirs. Consequently, as the capsules move through the tube, they sample discrete temperatures along the continuous gradient (Fig. \ref{Fig exp - thermal cycling}c). Since the capsules are nominally evenly spaced, each capsule experiences the same temperature profile over time, phase-shifted relative to its neighbors (faint traces compared to the bold trace in Fig. \ref{Fig exp - thermal cycling}c). The numbered events \cnum{1}-\cnum{5} correspond to different capsules in panel (a) simultaneously sampling different temperatures at distinct locations along the loop. When the capsules’ internal temperature exceeds a critical threshold near the reservoirs, they undergo rapid snap-through transitions (Fig. \ref{Fig exp - thermal cycling}d). As a result, the capsules alternate between open and closed states over time (Fig. \ref{Fig exp - thermal cycling}e), completing thermodynamic cycles that can be exploited for energy harvesting. Figs. \ref{Fig exp - thermal cycling}f-g show the typical length difference between the open and closed capsule states, recorded at locations \cnum{3} and \cnum{5} in the system, respectively, that corresponds to stored energy difference. 
These experiments capture the essential elements required for thermally driven capsule-based thermodynamic cycles. As an alternative means of accessing thermal variations, we next consider mobile systems and formulate a path optimization framework for operating across environmental thermal fields.

\subsection*{Energy-aware path optimization in a thermal field}\label{Subsection Path optimization algorithm}

\begin{figure}[h!]
\centerline{\includegraphics[width=1\textwidth]{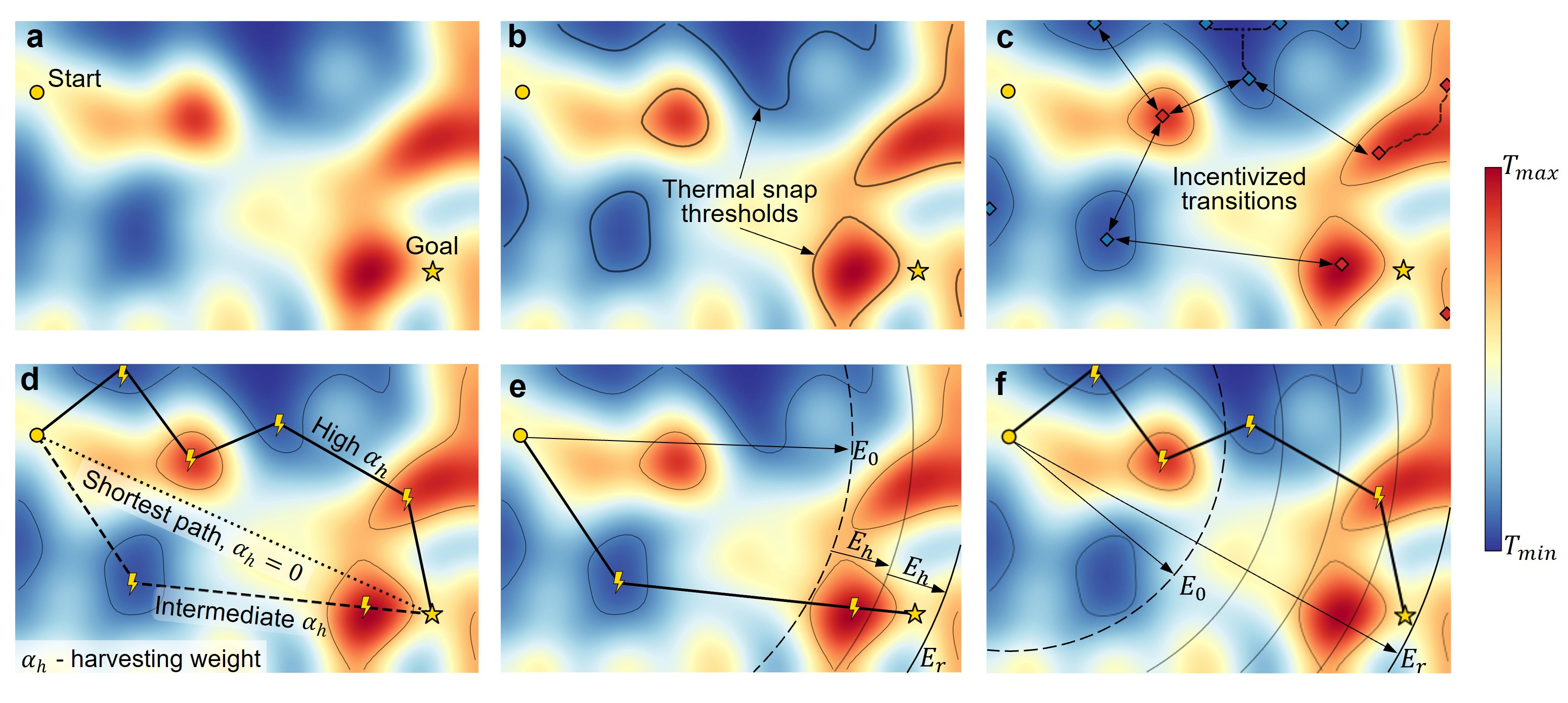}}
{\caption{\textbf{Optimal energy-aware path planning in a thermal field}. (a) Start and goal locations within a temperature field. (b) Thermal thresholds define boundaries that separate between hot and cold regions. Crossing these boundaries alternatingly triggers snap-through transitions in the multi-stable capsules. (c) Graph-based representation of the temperature field, with representative incentivized transitions between alternating thermal zones that enable energy harvesting. (d) Optimal simulated trajectories for increasing energy-harvesting incentives: the shortest path, a moderately incentivized path allowing limited detours, and a highly incentivized path favoring extended traversal through energy-harvesting regions. (e) A path requiring a moderate harvesting incentive to reach a target beyond the initial boundary range achievable with stored energy $E_{0}$, extending it through harvested energy $E_{h}$ at each snap-through event, yielding a total possible range $E_{r}$. (f) A path optimized with stronger harvesting incentive, achieving comparable range $E_{r}$, despite a smaller initial stored energy $E_{0}$ relative to (e), by traversing a greater number of energy-harvesting locations along the route.}
\label{Fig 5 - Optimal paths}}
\end{figure}

Since the capsules can be designed to respond to specified thermal conditions, navigation across thermal fields can replace fixed thermal reservoirs in performing thermodynamic cycles, making path planning a central component of the process. We develop an energy-aware path planning algorithm that computes optimal routes over heterogeneous temperature fields for a robot moving at constant speed (Fig. \ref{Fig 5 - Optimal paths}, Movie S4). The optimization balances locomotion cost with access to energy-harvesting zones, yielding trajectories that traverse alternating hot and cold areas that induce snap-through events in the embedded multistable capsules and enable sustained energy harvesting.

We partition the continuous thermal field into three regions defined by the snap temperature thresholds  $T_{open}$ and $T_{close}$, derived from the coupled elastic force-displacement response and internal gas pressure (see SI \S1.2). The corresponding zone indicator function is

\begin{equation}
Z(x, y) =
\left\{
\begin{array}{lll}
0,\quad  & T(x,y) \le T_{close}, & \text{(Cold)} \\[6pt]
1, & T_{close} < T(x,y) < T_{open}, \qquad &  \text{(Neutral)} \\[6pt]
2, & T(x,y) \ge T_{open}. & \text{(Hot)}
\end{array}
\right.
\label{zones}
\end{equation}

\vspace{5pt} 
   
The boundaries of the hot and cold regions define where the capsules undergo snap-through transitions (Fig. \ref{Fig 5 - Optimal paths}b). Next, we abstract this zonal map into a topology-preserving graph $G=(V,E)$, where each node $i\in V$ encodes spatial coordinates and the thermal zone label $Z_{i}$.
 
Zones below a minimal area threshold are represented by a single vertex, while larger areas are skeletonized for a reduced geometric representation (see SI \S3). To account for the energy cost of locomotion, each edge $(i,j)\in E$ with Euclidean length $d_{ij}$ is assigned a proportional energy cost $E_{move,ij}=\gamma d_{ij}$,
where $\gamma>0$ is the specific energy cost per unit distance.

Energy harvesting arises from snap-through transitions, with the harvested energy governed  by the capsule energy landscape, where $\Delta E_{1}$ and $\Delta E_{2}$ denote the energy released during snap-down and snap-up events, respectively (see SI \S2.5). The energy harvested along an edge is $E_{h,ij} = n_{c}\eta \Delta E_{k},\, \Delta E_{k\in\{1,2\}}\ge0$, where $n_{c}$ is the number of embedded capsules and $\eta$ is the conversion efficiency (SI \S2.3.3, Eq. 44). For clarity, we present results with a uniform energy gain per harvesting event $E_{h,ij}=E_{h}$, although generally the harvested energy can be tuned by modifying the capsule's energy profile.

In contrast to path planners that consider energy replenishment independent of the path taken \cite{plonski2013energy,hong2018range}, our algorithm accounts for the the capsules’ state, assigning edge weights that incentivize paths through alternating hot and cold thermal zones that induce snap-through and energy harvesting.
We therefore define an indicator function $\zeta(Z_{i}, Z_{j})$, that activates energy gain only for alternating hot-cold crossings (see Fig. \ref{Fig 5 - Optimal paths}c for representative $\zeta=1$ transitions),
\begin{equation}
    \zeta(Z_{i}, Z_{j}) =
\begin{cases}
1, & (Z_i, Z_j) \in \{(0,2), (2,0)\}, \\
0, & \text{otherwise}.
\end{cases}
\end{equation}
\vspace{3pt}

The optimal route $P^{*}$ is obtained by minimizing the net energy cost across all feasible paths $\mathcal{P}(s\rightarrow g)$ connecting the start and goal nodes,

\begin{equation}\label{opt_obj}
P^{*} = \arg \!\!\!\!\!\!\! \min_{P\in\mathcal{P}(s\rightarrow g)} \sum_{(i,j) \in P}
 E_{move,ij} \left(1 - \alpha_{h} \, \zeta(Z_{i}, Z_{j}) \cfrac{E_{h,ij}}{E_{move,ij}}\right),
\end{equation}

where the first term penalizes motion cost and the second term rewards energy harvested from thermal transitions. The parameter $\alpha_{h} \in [0,1]$ in~\eqref{opt_obj} controls the relative weighting between travel cost and energy harvesting, reflecting the relative reliance on intrinsic battery power versus energy acquisition. The formal definition of problem can be found in SI \S3, Eq. 68, which is solved using Dijkstra’s shortest-path algorithm \cite{dijkstra2022note}. 
While $\alpha_{h}=0$ yields the shortest path, an established solution in the literature \cite{gallo1988shortest}, we present solutions for $\alpha_{h}>0$, for which increasing $\alpha_{h}$ biases the planner toward energy-harvesting detours that cross hot-cold transitions (Fig. \ref{Fig 5 - Optimal paths}d). Thus, $\alpha_{h}$ reflects mission priorities, where low values are suitable when the destination lies within the intrinsic energy capacity, whereas higher values are essential to enable range extension beyond the stored energy limits through active energy harvesting.

Our path optimization results show that the operational range can be extended beyond the limit imposed by the initial onboard energy $E_{0}$ to a total energy $E_{r}$ through discrete energy harvesting gains (Fig. \ref{Fig 5 - Optimal paths}e). This extension is determined by the cumulative energy acquired along the trajectory, $E_{r}-E_{0}=N_{h}(\alpha_{h})E_{h}$, where $N_{h}(\alpha_{h})$ is the number of harvesting events along the path. 
Furthermore, prioritizing energy harvesting can achieve comparable range with lower initial stored energy (Fig. \ref{Fig 5 - Optimal paths}f), reducing the required onboard energy storage. The corresponding reduction in required stored energy between two paths is, $E_{0,II} - E_{0,I} =|E_{h} \left( N_{h} \left(\alpha_{h,II}\right) - N_{h} \left(\alpha_{h,I}\right) \right)|$, where $I$ and $II$ denote two distinct harvesting incentives. The reduction is illustrated in Fig. \ref{Fig 5 - Optimal paths}f, relative to the higher initial energy required in Fig. \ref{Fig 5 - Optimal paths}e. 

Overall, our formulation links the system’s environment, energy-harvesting mechanism, and navigational strategy, enabling motion-driven thermodynamic cycles. This framework can generalize to other spatially varying or dynamic fields capable of inducing snap-through transitions in the embedded energy units.

\section*{Discussion}
\label{Sec_Discussion}

Our study shows that thermodynamic cycles can emerge through active traversal of environmental thermal fields, reframing the environment as spatially distributed reservoirs accessed through mobility, replacing classical fixed reservoirs. These cycles are performed using embedded multistable unit cells that store energy across discrete stable states. Temperature variations drive states transitions, releasing energy through rapid instabilities that are harnessed to extract energy that would otherwise be dissipated.

We provide an analytical framework that captures the coupled dynamics and thermodynamics of the metafluid for embedded energy storage, harvesting, and release. This framework also characterizes the heat exchange mechanisms, that can extend to multicaloric composite systems \cite{multicaloric_cooling_hou2022materials}, where coupled thermal, mechanical, and electromagnetic stimuli govern heat exchange, holding great potential for improving cooling performance.
The thermodynamic efficiencies of the energy storage processes presented in Figs. \ref{Fig 3 - Energy storage and work}-\ref{Fig exp - thermal cycling} are $86.54\%$, $28.67\%$, and $66.81\%$, respectively (see SI \S2.6). Future work will focus on optimizing the system design and dynamic operation to further improve thermodynamic efficiency. 

Another direction is to investigate and optimize the electrical energy extraction from the rapid snap-through motion, including design and positioning of the electromagnets, as well as high-temporal-resolution characterization of the transient response.
Alternative transduction mechanisms, including piezoelectric \cite{mahapatra2021piezoelectric}, triboelectric \cite{wu2019triboelectric,cheng2023triboelectric}, magnetostrictive \cite{deng2018review}, capacitive transduction \cite{yen2006variable}, or combination thereof through hybrid energy harvesters \cite{ryu2019hybrid}, could be explored as mechanisms for efficient conversion of the energy associated with the rapid snap-through motion.

More generally, the proposed system belongs to a class of out-of-equilibrium systems that exhibits multiscale dynamics across length and time scales, with gradual shifts driven by external inputs until stability thresholds are crossed, leading to rapid release through dynamical instability. This mechanism appears across diverse physical domains, suggesting further implementations leveraging phase transitions \cite{li2021phase_transitions}, stick-slip dynamics \cite{stick_slip1994dynamic}, and magnetic domain switching \cite{manchon2019current}. This behavior enables programmability across scales \cite{responsive_metamaterials2024}, from the design of the energy profile of individual units to navigation strategies through environmental gradients, or conversely, inverse design tailored to specific environments.

This work opens a broader research direction at the intersection of thermodynamics and robotics, in which thermodynamic state variables become dynamic objectives within a robot’s decision-making framework. It raises exciting new research questions on optimal foraging \cite{pyke1984optimal} in autonomous energy-seeking systems and on the efficiency limits of cycles that harvest environmental energy while powering locomotion. Future research will investigate energy-aware trajectory planning based on local information in environments where temperature distributions vary in time or are only partially known, while accounting for the robotic system needs to have a sufficient initial energy budget to sense and traverse thermal gradients. In this perspective, thermodynamic state shifts from a background parameter to a governing driver of behavior, particularly under conditions of energy scarcity.

\section*{Methods}\label{Methods}

\subsection*{Analytical model, numerical simulations and experimental details}\label{Methods - Analytical and numerical models}

All details of the analytical model, simulations, and experiments are provided in the Supplementary Information.

\section*{Data availability}
Data are publicly accessible at  \url{https://doi.org/10.6084/m9.figshare.31400337}. 
 
\section*{Code availability}
All code supporting this study is available online at \url{https://doi.org/10.6084/m9.figshare.31400337}.

\section*{Acknowledgments}
S.K. acknowledges the support of the Clore Scholars Programme of the Clore Foundation. E.B.A. is grateful to the Azrieli Foundation for the award of an Azrieli Fellowship.

\section*{Author contributions}
S.K., S.G., and A.D.G. conceived the research idea and developed the analytical model. S.K. designed and performed the experiments, analyzed the data, and developed the numerical tools. E.B.A. fabricated the experimental thermal cycling system and supported experiments. Y.V. supported the optimization framework. S.K., A.Z., and A.D.G. wrote and edited the manuscript and the supplementary information. S.G. and A.D.G. supervised the project. All authors contributed to the final manuscript.

\section*{Competing interests}
The authors declare no competing interests.

\section*{Additional information}
\backmatter

\bmhead{Supplementary information}
The online version contains supplementary material.

\bmhead{Correspondence}
and requests for materials should be addressed to Sofia Kuperman.

\bibliography{Bibliography}

\end{document}